\documentclass[preprint,aps]{revtex4}
\usepackage{graphicx}

\begin{document}




\title{Selective Hybridization between Main Band and Superstructure Band in Bi$_2$Sr$_2$CaCu$_2$O$_{8+\delta}$ Superconductor}

\author{Qiang Gao$^{1,2}$, Hongtao Yan$^{1,2}$, Jing Liu$^{1,2,3}$, Ping Ai$^{1,2}$, Yongqing Cai$^{1,2}$, Cong Li$^{1,2}$, Xiangyu Luo$^{1,2}$, Cheng Hu$^{1,2}$, Chunyao Song$^{1,2}$, Jianwei Huang$^{1,2}$, Hongtao Rong$^{1,2}$, Yuan Huang$^{1}$, Qingyan Wang$^{1}$, Guodong Liu$^{1,2}$, Genda Gu$^{4}$, Fengfeng Zhang$^{5}$, Feng Yang$^{5}$, Shenjin Zhang$^{5}$, Qinjun Peng$^{5}$, Zuyan Xu$^{5}$, Lin Zhao$^{1}$, Tao Xiang$^{1,2,3,6}$ and X. J. Zhou$^{1,2,3,6,*}$}
\affiliation{
\\$^{1}$Beijing National Laboratory for Condensed Matter Physics, Institute of Physics, Chinese Academy of Sciences, Beijing 100190, China.
\\$^{2}$University of Chinese Academy of Sciences, Beijing 100049, China.
\\$^{3}$Beijing Academy of Quantum Information Sciences, Beijing 100193, China
\\$^{4}$Condensed Matter Physics and Materials Science Department, Brookhaven National Laboratory, Upton, New York, 11973, USA
\\$^{5}$Technical Institute of Physics and Chemistry, Chinese Academy of Sciences, Beijing 100190, China.
\\$^{6}$Songshan Lake Materials Laboratory, Dongguan, Guangdong 523808, China.
\\$^{*}$Corresponding authors: XJZhou@iphy.ac.cn
}

\date{\today}

\begin{abstract}
\bf High-resolution laser-based angle-resolved photoemission measurements have been carried out on Bi$_2$Sr$_2$CaCu$_2$O$_{8+\delta}$ (Bi2212) and Bi$_2$Sr$_{2-x}$La$_x$CuO$_{6+\delta}$ (Bi2201) superconductors. Unexpected  hybridization between the main band and the superstructure band in Bi2212 is clearly revealed. In the momentum space where one main Fermi surface intersects with one superstructure Fermi surface, four bands are observed instead of two. The hybridization exists in both superconducting state and normal state, and in Bi2212 samples with different doping levels. Such a hybridization is not observed in Bi2201. This phenomenon can be understood by considering the bilayer splitting in Bi2212, the selective hybridization of two bands with peculiar combinations, and the altered matrix element effects of the hybridized bands. These observations provide strong evidence on the origin of the superstructure band which is intrinsic to the CuO$_2$ planes. Therefore, understanding physical properties and superconductivity mechanism in Bi2212 should consider the complete Fermi surface topology which involves the main bands, the superstructure bands and their interactions.
\end{abstract}
\maketitle

\newpage

High temperature cuprate superconductors have been extensively studied for more than thirty years due to its unusally high critical temperature ($T_C$), anomalous normal state, and challenging mechanism of high temperature superconductivity \cite{JRKirtley2000CCTsuei,BStatt1999TTimusk,ZXShen2003ADamascelli,XGWen2006PALee,JZaanen2015BKeimer}. Angle-resolved photoemission spectroscopy (ARPES) \cite{ZXShen2003ADamascelli,JCCampuzano2004,XJZhou2007} has played a key role in studying the electronic structure of the cuprate superconductors, including the revelation of the distinct d-wave superconducting gap symmetry \cite{CHPark1993ZXShen,HDing1996,ZXShen2014MHashimoto}, the pseudogap above $T_C$ \cite{ZXShen1996DSMarshall,KKapitulnik1996AGLoeser,JGiapintzakis1996HDing,ZXShen2018IMVishik} and many-body effects \cite{ZXShen2000PVBogdanov,DGHinks2001PDJohnson,KKadowaki2001AKaminski,ZXShen2001ALanzara,ZXShen2003XJZhou,DSDessau2003ADGromko,JFink2003TKKim,ZXShen2004TCuk,XJZhou2008WTZhang,XJZhou2013JFHe,XJZhou2016JMBok,ZXShen2018YHe}. 
The majority of these significant ARPES results are obtained from Bi$_2$Sr$_2$CaCu$_2$O$_{8+\delta}$ (Bi2212) because of the availability of high quality single crystals and its easiness to cleave to get smooth and clean surfaces. Bi2212 is also widely used in other experimental techniques like scanning tunneling microscope/spectroscopy (STM/STS) \cite{JCDavis2001SHPan,JCDavis2002JHoffman,JCDavis2007Yohsaka,JCDavis2016MHamidian,OFisher2007}.

It is well-known that the bismuth-based cuprate superconductors have incommensurate modulations in their crystal structure along the $b$* direction \cite{AWSLEIGHT1988YGao,BGHyde1988RLWithers,OEibl1991,LJGauckler1994HHeinrich}. This modulation leads to the formation of superstructure bands of various orders in the measured ARPES results \cite{KKadowaki1994PAebi,KKadowaki1995JOsterwalder,GPBrivio1996HDing,KKadowaki2000HMFretwell}. A related issue under debate is about the origin of the superstructure bands whether they are extrinsic due to diffraction of the BiO layers, or intrinsic coming directly from the CuO$_2$ planes \cite{KKadowaki1995JOsterwalder,GPBrivio1996HDing,XJZhou2019JLiu}. Another feature of Bi2212 is that there are two CuO$_2$ planes in one structural unit separated by calcium (Ca). The interaction between these two structurally equivalent CuO$_2$ planes gives rise to bilayer splitting, i.e., two Fermi surface sheets, bonding and antibonding, which correspond to different doping levels with distinct superconducting gap \cite{ZXShen2001PVBogdanov,ZXShen2001DLFeng,DSDessau2001YDChuang,
RFollath2002SVBorisenko,TXiang2003YHSu,STurchini2004SVBorisenko,DSDessau2004YDChuang,
YAndo2004AAKordyuk,XJZhou2019PAi}. While most ARPES measurements focus on the main band, little attention has been paid on the interaction between the main and superstructure bands, and between the bonding and antibonding bands. These interactions may provide important information on the origin of superstructure bands and adequate consideration of Fermi surface topology in studying the physical properties and superconductivity mechanism in Bi2212.

In this paper, we report the observation of an unexpected hybridization between the main band and superstructure band in Bi2212 due to the interaction between the bonding and antibonding bands. Taking advantage of high resolution laser-based ARPES measurements, we observed four bands, instead of two, in the momentum space where one main band intersects with one superstructure band. Such an unusual hybridization is present in both the normal state and the superconducting state and in Bi2212 samples with different doping levels. This hybridization phenomenon is not observed in the similar ARPES measurements on Bi$_2$Sr$_{2-x}$La$_x$CuO$_{6+\delta}$ (Bi2201) excluding the possibility of its spin-related origin. We propose this unusual band hybridization in Bi2212 can be understood by considering the bilayer splitting effect. We find that in this case only the bonding and antibonding bands can hybridize and the resultant hybridized bands exhibit distinct matrix element effects. These observations provide key insights on the origin of the superstructure band that is intrinsic to the CuO$_2$ planes. They also provide a new overall picture of the Fermi surface topology of Bi2212 that needs to be taken into account in studying its electronic structure and physical properties.
 

ARPES measurements were carried out on a vacuum ultraviolet (VUV) laser-based ARPES system \cite{XJZhou2008GDLiu,XJZhou2018}. The photon energy is 6.994 eV with a bandwidth of 0.26 meV. The energy resolution of the electron energy analyzer (Scienta DA30L) was set at 1 meV, giving rise to an overall energy resolution of 1.0 meV. The angular resolution was $\sim$0.3$^\circ $, corresponding to a  momentum resolution of 0.004 $\mathring{A}^{-1}$. The Fermi level is referenced by measuring on the Fermi edge of a clean polycrystalline gold that is electrically connected to the sample. High quality optimally-doped Bi$_2$Sr$_{2-x}$La$_x$CuO$_{6+\delta}$ (x=0.4, $T_C$=32 K, denoted as OP32K hereafter) \cite{XJZhou2009JQMeng} and Bi$_2$Sr$_2$CaCu$_2$O$_{8+\delta}$ ($T_C$=91 K, denoted as OP91K hereafter) single crystals were grown by the floating zone method. Overdoped Bi2212 sample ($T_C$=78 K, denoted as OD78K hereafter) was prepared by annealing the optimally-doped sample in high pressure oxygen atmosphere \cite{XJZhou2016YXZhang}. Underdoped Bi2212 sample ($T_C$=64 K, denoted as UD64K hereafter) was prepared by annealing the optimally-doped sample in vacuum. The $T_C$ was measured using a Quantum Design SQUID magnetometer. All samples were cleaved $in$ $situ$ and measured in vacuum with a base pressure better than $3\times10^{-11}$ mbar.

\begin{figure*}[tbp]
\begin{center}
\includegraphics [width=1.0\columnwidth,angle=0]{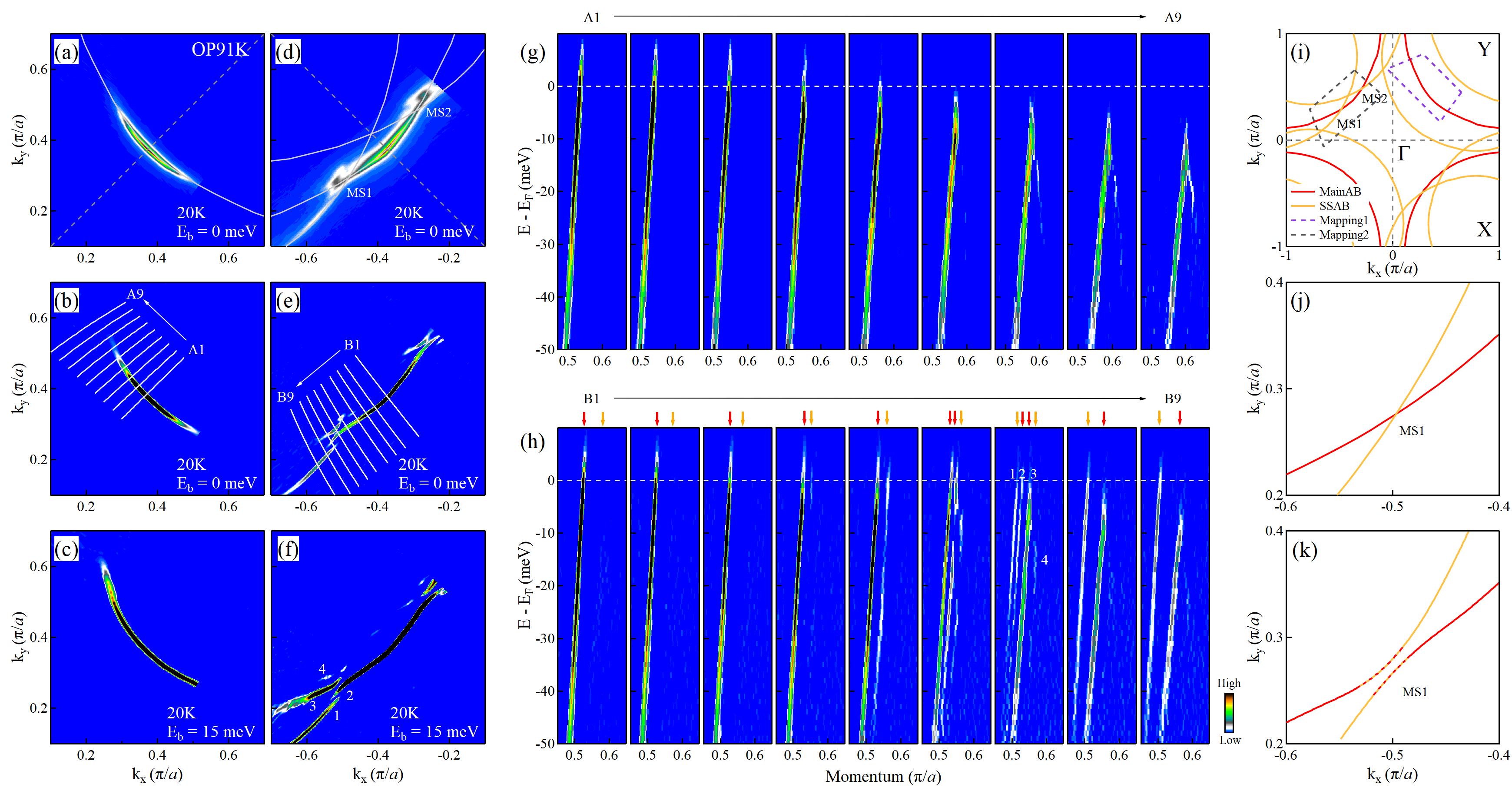}
\end{center}
\caption { \textbf{Fermi surface and band structure of Bi2212 (OP91K) measured at a temperature of 20 K by laser-based ARPES.} (a) Fermi surface mapping in the first quadrant of the Brillouin zone. It is obtained by integrating the spectral weight over an energy window of [-2,+2] meV with respect to the Fermi level. (b) The second derivative of the original image (a) with respect to momentum. The second derivative image helps enhance the contrast to reveal some detailed structures. (c) The second derivative image of the constant energy contour obtained by integrating the spectral weight over an energy window of [-3,+3] meV with respect to a binding energy of 15 meV. (d-f) The same as (a-c) but measured in the second quadrant of the Brillouin zone. (g) Band structure along different momentum cuts A1-A9 in the first quadrant. The location of the momentum cuts is shown by white lines in (b). The images are second derivative of the original band structure with respect to momentum. (h) Band structure along different momentum cuts B1-B9 in the second quadrant. The location of the momentum cuts is shown by white lines in (e). The images are second derivative of the original band structure with respect to momentum. The red and yellow arrows mark the location of the observed bands. (i) Schematic Fermi surface of Bi2212 containing the main antibonding Fermi surface (MainAB, red lines) and  its first-order superstructure replicas (SSAB, yellow lines). The dashed purple box represents the mapping area of (a), and the dashed grey box represents the mapping area of (d). (j) The enlarged momentum area where the main band and the superstructure band intersects at the MS1 point. (k) Same as (j) by assuming hybridization of two bands. 
}
\label{Fig1}
\end{figure*}

Figure \ref{Fig1} shows the Fermi surface and the corresponding band structure of the Bi2212 OP91K sample. Two momentum spaces are covered, one in the first quadrant and the other in the second quadrant, as indicated in Fig. \ref{Fig1}(i) by the dashed box areas. The Fermi surface measured in the first quadrant (Fig. \ref{Fig1}(a) and Fig. \ref{Fig1}(b)) consists of only one sheet, as also shown by the measured band structures in Fig. \ref{Fig1}(g). This Fermi surface corresponds to the antibonding sheet when measured by using 6.994 eV laser \cite{XJZhou2008WTZhang}; the bonding sheet is fully suppressed at this measurement condition. In the second quadrant, the main Fermi surface is expected to intersect with two first-order superstructure replicas at MS1 and MS2 points, as indicated in Fig. \ref{Fig1}(i). However, the measured Fermi surface in the second quadrant (Figs. \ref{Fig1}(d), (e)) shows rather complex topology near the crossing areas, MS1 and MS2, of the main and superstructure Fermi surface sheets. First, both the main Fermi surface and the superstructure Fermi surface get broken and become disconnected near the MS1 crossing point. Three branches of Fermi surface can be clearly seen near the crossing point although the main Fermi surface and the superstructure Fermi surface still consist of only one band away from the crossing point. In the constant energy contour at a binding energy of 15 meV (Fig. \ref{Fig1}(f)), these features show up more clearly by reducing the effect of the superconducting gap and four branches of  contours can be identified near the MS1 crossing point. Here the branch 1 comes from the superstructure band, and the branches 2 and 3 come from the main band broken at the crossing point. But the weak branch 4 appears unexpectedly which can not be attributed to either the main band or the superstructure band. 

When the momentum cuts move from the nodal direction to antinodal direction (B1 to B9 cuts in Fig. \ref{Fig1}(e)), the measured band structure (Fig. \ref{Fig1}(h)) changes from two bands for B1-B5 cuts, to three bands for B6 cut, to four bands for B7 cut, and finally back to two bands for the cuts B8 and B9, as indicated by the arrows on the top of Fig. \ref{Fig1}(h). The band structure evolution in Fig. \ref{Fig1}(h) agrees with the measured Fermi surface in Fig. \ref{Fig1}(e) and the constant energy contour in Fig. \ref{Fig1}(f). In particular, the simultaneous observation of four bands in Cut B7 corresponds to the four branches of contours in Fig. \ref{Fig1}(f). We note that  the band 4 in the Cut B7 panel of Fig. \ref{Fig1}(h) gives rise to the branch 4 in Fig. \ref{Fig1}(f); it is not due to the superconductivity-induced back bending of the band 3 because it is parallel to the band 3 \cite{TKondo2017SKunisada}. The bands caused by superconductivity-induced back bending exhibits different slope from the original band, as seen in the Cuts A7-A9 panels in Fig. \ref{Fig1}(g). The two bands observed away from the crossing point, for cuts B1-B5, B8 and B9, are well understood as from one main Fermi surface and one superstructure Fermi surface as depicted in Fig. \ref{Fig1}(d). However, the simultaneous observation of four branches of Fermi surface sheets (Fig. \ref{Fig1}(f)) and four bands (Fig. \ref{Fig1}(h) B7) near the crossing point MS1 is rather different from the expected pictures where only two bands can be observed, either there is no hybridization (Fig. \ref{Fig1}(j)) or hybridization (Fig. \ref{Fig1}(k)) between them.

\begin{figure*}[tbp]
\begin{center}
\includegraphics [width=1.0\columnwidth,angle=0]{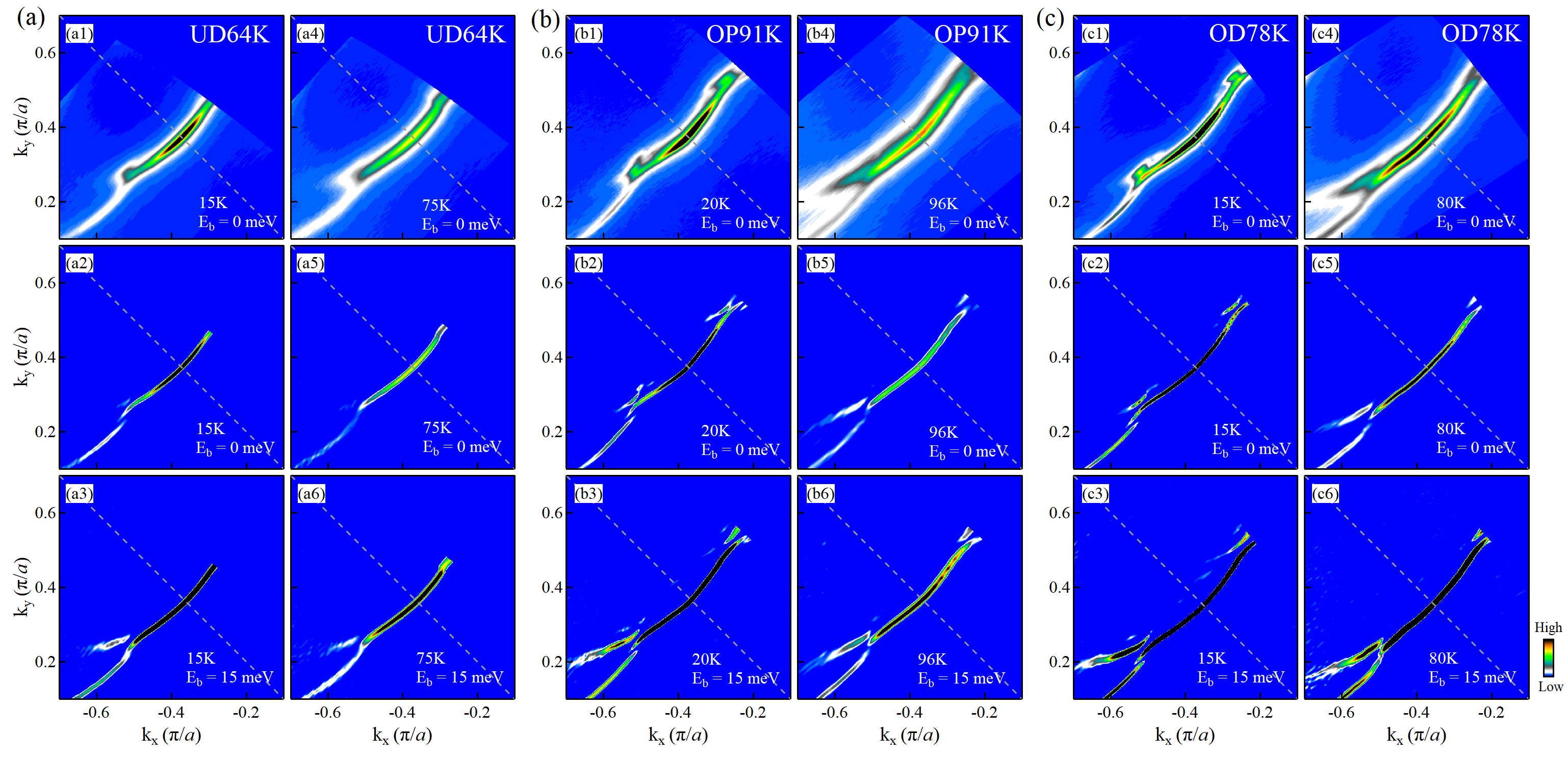}
\end{center}
\caption { \textbf{Doping and temperature dependences of Fermi surface and band hybridization in Bi2212.} (a) Fermi surface and constant energy contour of underdoped Bi2212 with $T_c$=64 K (UD64K) measured in the superconducting state at 15 K (a1-a3) and the normal state at 75 K (a4-a6). (a2), (a3), (a5) and (a6) are obtained by the second derivative of the original data with respect to momentum. (a1), (a2), (a4) and (a5) represent the measured Fermi surface while (a3) and (a6) are constant energy contours at a binding energy of 15 meV. (b) Same as (a) but the Fermi surface and constant energy contour are measured on an optimally-doped Bi2212 with $T_c$=91 K (OP91K) in the superconducting state (20 K) and the normal state (96 K). (c) Same as (a) but the Fermi surface and constant energy contour are measured on an overdoped Bi2212 with $T_c$=78 K (OD78K) in the superconducting state (15 K) and the normal state (80 K).  
}
\label{Fig2}
\end{figure*}

To check the effect of superconductivity on this unusual band hybridization and its doping dependence, we measured Bi2212 samples with different doping levels in both normal state and superconducting state. Fig. \ref{Fig2} shows the Fermi surface mappings and the constant energy contours at a binding energy of 15 meV for the underdoped UD64K (Fig. \ref{Fig2}(a)), optimally-doped OP91K (Fig. \ref{Fig2}(b)), and overdoped OD78K (Fig. \ref{Fig2}(c)) samples both above and below their corresponding  superconducting transition temperatures. The measured data in the normal state show quite similar behaviours as that measured in the superconducting state, i.e., the Fermi surface topology near the crossing point is characterized by broken Fermi surface and the appearance of multiple branches. Some fine structures, in particular the appearance of the branch 4, become less clear in the normal state data possibly due to thermal broadening at relatively high temperature. This will not affect the result that unusual band hybridization exists both above and below $T_C$ and therefore it is not caused by superconductivity. The three samples with different doping levels show quite similar behaviours near the crossing area of the main band and the superstructure band. Here some fine structures in the measured data of the underdoped UD64K sample (Fig. \ref{Fig2}(a)) are less clear than those in the optimally-doped and overdoped samples which may be related to photoemission spectrum broadening by strong electron scattering in the underdoped samples \cite{ZXShen2000DLFeng,ZXShen2004XJZhou,ZXShen2005KMShen,ZXShen2006KTanaka}. The key features of the unusual band hybridization, broken Fermi surfaces and multiple branches, remain in the underdoped sample. Therefore, the unusual band hybridization between the main and the superstructure bands is a general phenomenon in Bi2212 with different doping levels.

\begin{figure*}[tbp]
\begin{center}
\includegraphics [width=1.0\columnwidth,angle=0]{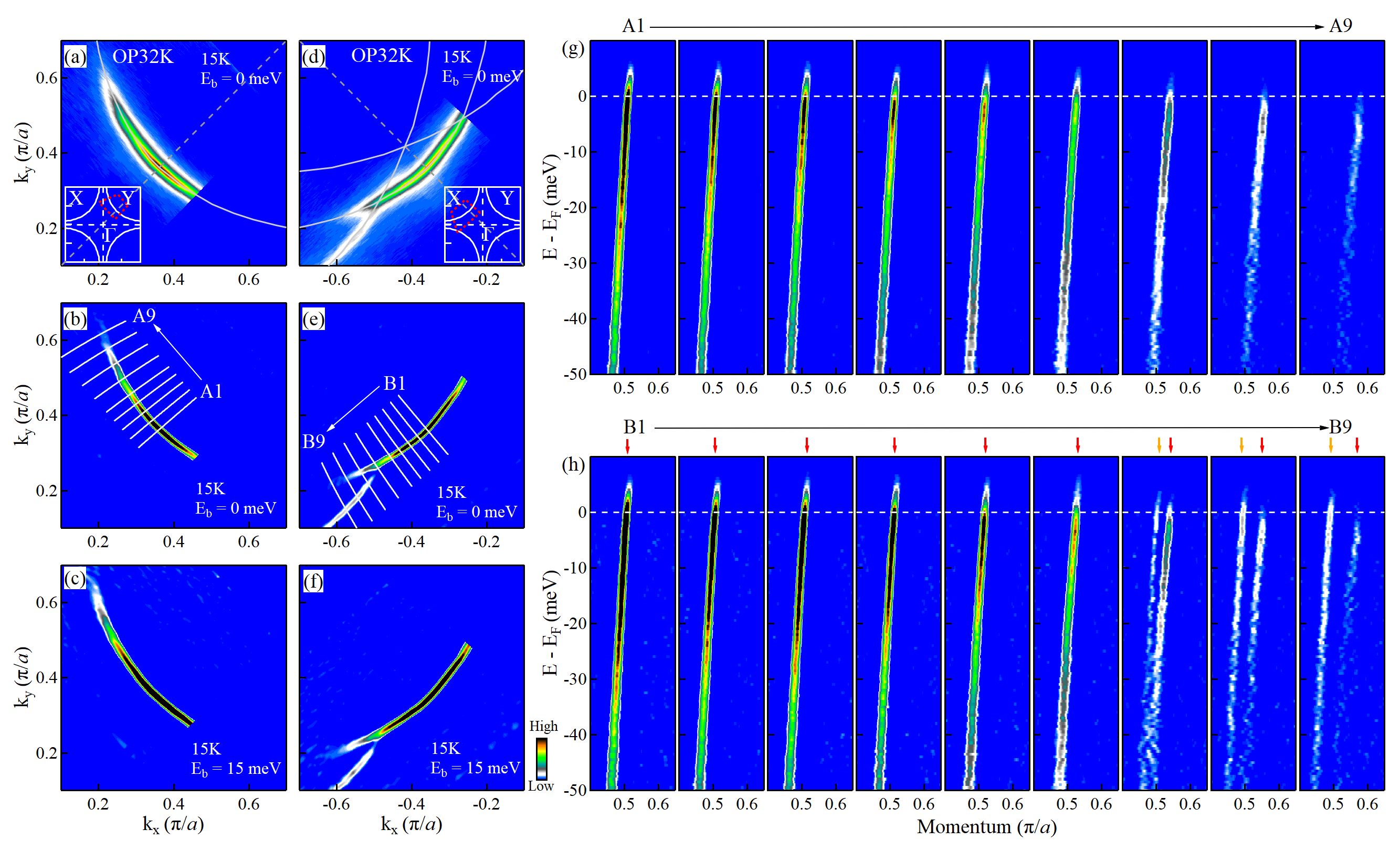}
\end{center}
\caption {\textbf{Fermi surface and band structure of Bi2201 (OP32K) measured at a temperature of 20 K by laser-based ARPES.} (a) Fermi surface mapping in the first quadrant of the Brillouin zone. It is obtained by integrating the spectral weight over an energy window of [-2,+2] meV with respect to the Fermi level. The corresponding momentum area is marked as the dashed red box in the inset. (b) The second derivative of the original data corresponding to (a) with respect to momentum. (c) Constant energy contour at a binding energy of 15 meV. The image is obtained by the second derivative of the original data with respect to momentum. (d) Fermi surface mapping in the second quadrant of the Brillouin zone at 15 K. The corresponding momentum area is marked as the dashed red box in the inset. (e) The second derivative of the original data corresponding to (d) with respect to momentum. (f) Constant energy contour at a binding energy of 15 meV. The image is obtained by the second derivative of the original data with respect to momentum. (g)-(h) Band structure measured at 15 K. The location of the corresponding momentum cuts is shown as white lines in (b) and (e) labelled with A1-A9 and B1-B9. The images are obtained by the second derivative of the original data with respect to momentum. The red and yellow arrows in (h) mark the location of the observed bands.}
\label{Fig3}
\end{figure*}

Since superstructure modulation is present in all the bismuth-based superconductors \cite{AWSLEIGHT1988YGao,BGHyde1988RLWithers,OEibl1991,LJGauckler1994HHeinrich}, it is natural to ask whether similar band hybridization phenomenon can also occur in other systems besides Bi2212. To this end, we performed similar ARPES measurements on Bi2201. The measured Fermi surface in two quadrants and the corresponding band structure are shown in Fig. \ref{Fig3}. In the first quadrant, single Fermi surface sheet is clearly observed (Fig. \ref{Fig3}(a)). The corresponding band structure consists of a single band for the momentum cuts from nodal to antinodal regions (Fig. \ref{Fig3}(g)). In the second quadrant, both the main Fermi surface and the superstructure replica are clearly observed. The main Fermi surface is continuous and there is no additional Fermi surface branch appeared in the crossing area of the main and superstructure bands (Fig. \ref{Fig3}(d)). There appear only the main band and the superstructure band in the measured band structure for the momentum cuts from the nodal to the antinodal directions and no sign of any additional bands is observed (Fig. \ref{Fig3}(h)). These results can be well understood by the picture in Fig. \ref{Fig1}(j) where the main band and the superstructure band cross without hybridization. The unusual band hybridization observed in Bi2212 does not occur in Bi2201 although in both cases only one main Fermi surface is observed.

Now we come to discuss on the possible origin of the peculiar band hybridization discovered in Bi2212. When two bands hybridize with each other, they usually still give rise to two avoided bands as shown in Fig. \ref{Fig1}(k). When we first observed two hybridized bands to split into four bands, the first possibility coming into mind is whether this is spin related phenomenon. Special momentum-locked spin texture has been reported in Bi2212 superconductor \cite{ALanzara2018KGotlieb}. If spin-up and spin-down bands split during the band hybridization between the main band and the superstructure band, it will naturally produce  four bands as we have observed. We believe this scenario is less likely. First, it remains to see whether the observed spin texture in Bi2212 can cause the splitting of the spin-up and spin-down bands. Second, if we assume the spin texture observed in Bi2212 is general in cuprate superconductors, the absence of the unusual band hybridization in Bi2201 is not consistent with the spin-related  scenario.

\begin{figure*}[tbp]
\begin{center}
\includegraphics [width=1.0\columnwidth,angle=0]{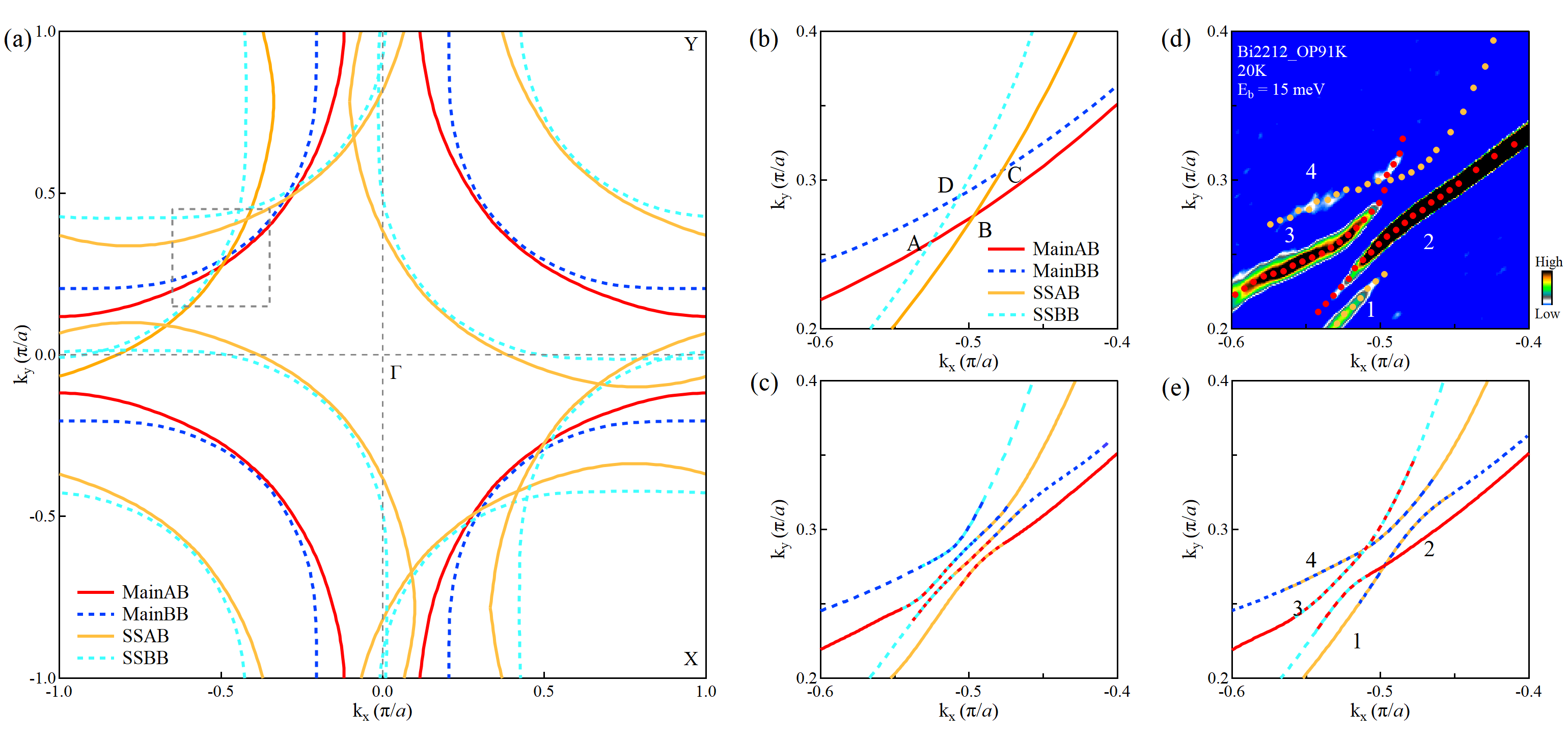}
\end{center}
\caption {\textbf{Schematic illustration of band hybridization and its comparison with the measured results in Bi2212.} (a) Schematic Fermi surface of Bi2212 containing the main Fermi surface with bilayer splitting: antibonding sheet (MainAB, red lines) and bonding sheet (MainBB, dashed blue lines), and first-order superstructure replicas: antibonding sheet (SSAB, yellow lines) and bonding sheet (SSBB, dashed green lines). The dashed grey box marks the momentum area enlarged in (b)-(e). (b) The enlarged momentum area marked in (a). The main bonding and antibonding Fermi surface sheets intersect with the superstructure bonding and antibonding replicas that gives rise to four crossing points labelled as A, B, C and D. (c) Same as (b) but considering band hybridization at all four crossing points. (d) The measured constant energy contour at a binding energy of 15 meV at the crossing point MS1 as shown in Fig.~\ref{Fig1}(f). The observed four branches of sheets are marked. (e) Same as (b) but considering the band hybridization only at A and C crossing points.
}


\label{Fig4}
\end{figure*}

The observation of the unusual band hybridization in Bi2212 and its absence in Bi2201 indicates that it is a Bi2212 specific phenomenon. Compared with Bi2201, the unique features of Bi2212 lie in the presence of two CuO$_2$ planes in one structural unit that gives rise to bilayer splitting \cite{ZXShen2001PVBogdanov,ZXShen2001DLFeng,DSDessau2001YDChuang,
RFollath2002SVBorisenko,TXiang2003YHSu,STurchini2004SVBorisenko,DSDessau2004YDChuang,
YAndo2004AAKordyuk,XJZhou2019PAi}. Although there is only one main Fermi surface that is observed during our measurements, there are actually two Fermi surface sheets, bonding and antibonding, that are present. This prompted us to examine whether this unusual hybridization could be related to the bilayer splitting in Bi2212. Fig. \ref{Fig4}(a) shows the Fermi surface topology of Bi2212 by considering both the bilayer splitting and the superstructure bands. In the same momentum area as covered by our experiment, two main Fermi surface sheets are expected to cross with two superstructure Fermi surface sheets in the second quadrant, giving rise to four crossing points (A, B, C and D) as shown in Fig. \ref{Fig4}(b). If there is no band hybridization at the four crossing points, it is obviously not consistent with the measured result shown in Fig. \ref{Fig4}(d). When we consider band hybridization on all the four crossing points in Fig. \ref{Fig4}(c), it does not agree with the measured result in Fig. \ref{Fig4}(d) either. A consistent picture emerges when we consider (1) only the bonding band and the antibonding band can hybridize. This applies to the crossing point A where the main antibonding band (MainAB) intersects with the superstructure bonding band (SSBB) and the crossing point C where the main bonding band (MainBB) intersects with the superstructure antibonding band (SSAB); (2) The bonding band does not hybridize with the bonding band like the crossing point D; The antibonding band does not hybridize with the antibonding band like the crossing point B; (3) In our present ARPES measurement condition, only the antibonding band is observed while the bonding band is fully suppressed. The hybridized bands show up because they are mixed with both bonding and antibonding characters. The resultant Fermi surface picture (Fig. \ref{Fig4}(e)) agrees well with the observed results (Fig. \ref{Fig4}(d)). In particular, the simultaneous observation of four bands in the crossing area can be well understood in this picture. 

The present work provides a unique opportunity to look into the interaction between the bonding and the antibonding bands in Bi2212. Our results indicate that only the bonding and the antibonding bands can hybridize with each other. The underlying mechanism of this selective hybridization is interesting that asks for further theoretical understanding. This band interaction rule can also explain why there is no band hybridization between the main and the superstructure bands in Bi2201. 

It has been under debate on the origin of superstructure bands in the bismuth-based superconductors in which the superstructure modulation is present \cite{KKadowaki1995JOsterwalder,GPBrivio1996HDing,XJZhou2019JLiu}. In one scenario, the superstructure bands are extrinsic that are formed when the photoelectrons from the CuO$_2$ plane are diffracted in passing through the BiO layer with superstructure modulation. In the other scenario, they are intrinsic that come from the CuO$_2$ planes directly. Our present work provides strong evidence in favour of their intrinsic nature. If the superstructure band originates from photoelectron diffraction, it is impossible for it to hybridize with the main band. Both the main band and the superstructure band, as well as their hybridization, come directly from the CuO$_2$ plane. This is consistent with the observation of superstructure modulation in CuO$_2$ planes \cite{JQLi2017CGuo,XJZhou2019JLiu}. The intrinsic nature of the main bands, the superstructure bands and their hybridization provides a complete Fermi surface topology for Bi2212 that should be considered in understanding its electron structure, physical properties and high temperature superconductivity mechanism.


In summary, by taking high resolution laser-based ARPES measurements on Bi2212, we have discovered an unexpected band hybridization between the main band and the superstructure band. Such a hybridization is observed in Bi2212 with different dopings and in both the normal and the superconducting states. It is not present in Bi2201. This phenomenon can be understood by considering the bilayer splitting and the selective interaction of bonding and antibonding bands in Bi2212. Our results provide strong evidence to support the intrinsic nature of the superstructure bands. They also provide a complete Fermi surface topology of Bi2212 that needs to be considered in theoretical and experimental understanding of its physical properties.


This work is supported by the National Key Research and Development Program of China (Grant No.2016YFA0300300 and 2017YFA0302900), the National Natural Science Foundation of China (Grant No. 11888101), the Strategic Priority Research Program (B) of the Chinese Academy of Sciences (XDB25000000), the Research Program of Beijing Academy of Quantum Information Sciences (Grant No. Y18G06), National Science Foundation of China (Grant No. 11874405), and the National Key Research and Development Program of China (Grant No. 2019YFA0308000). The work at Brookhaven was supported by the Office of Basic Energy Sciences, U.S. Department of Energy (DOE) under Contract No de-sc0012704.

X.J.Z., T.X. and Q.G. proposed and designed the research. G.D.G., Q.G. and H.T.Y. prepared single crystal. Q.G. carried out the experiment with H.T.Y., J.L. and P.A.; Q.G., H.T.Y., J.L., P.A., Y.Q., C.L., X.Y.L., C.H., C.Y.S., J.W.H., H.T.R., L.Z., Y.H., Q.Y.W., G.D.L., F.F.Z., F.Y., Q.J.P., Z.Y.X. and X.J.Z. contributed to the development and maintenance of Laser ARPES system. Q.G. and X.J.Z. analyzed the data. X.J.Z., Q.G. and L.Z. wrote the paper. All authors discussed the results and commented on the manuscript.

\newpage
\bibliographystyle{unsrt}

\begin{thebibliography}{10}

\bibitem{JRKirtley2000CCTsuei}
C. C. Tsuei and J. R. Kirtley,
\newblock Pairing symmetry in cuprate superconductors,
\newblock {Rev. Mod. Phys.} {\bf 72}, 969 (2000).

\bibitem{BStatt1999TTimusk}
T. Timusk and B. Statt,
\newblock The pseudogap in high-temperature superconductors: an experimental survey,
\newblock {Rep. Prog. Phys.} \textbf{62}, 61 (1999).

\bibitem{ZXShen2003ADamascelli}
A. Damascelli, Z. Hussain and Z. X. Shen,
\newblock Angle-resolved photoemission studies of the cuprate superconductors,
\newblock {Rev. Mod. Phys.} \textbf{75}, 473 (2003).

\bibitem{XGWen2006PALee}
P. A. Lee, N. Nagaosa, and X. G. Wen,
\newblock Doping a mott insulator: Physics of high-temperature superconductivity,
\newblock {Rev. Mod. Phys.} \textbf{78}, 17 (2006).

\bibitem{JZaanen2015BKeimer}
B. Keimer, S. A. Kivelson, M. R. Norman, S. Uchida, and J. Zaanen,
\newblock From quantum matter to high-temperature superconductivity in copper oxides,
\newblock {Nature} \textbf{518}, 179 (2015).

\bibitem{JCCampuzano2004}
J. C. Campuzano {\em et al.},
\newblock in \textit{The physics of superconductors}, edited by K. H. Bennemann and J. B. Ketterson 
\newblock {(Springer, New York, 2004)} Vol. 2.

\bibitem{XJZhou2007}
X. J. Zhou {\em et al.},
\newblock in \textit{Hand book of high-temperature superconductivity: Theory and Experiment}, edited by J.R. Schrieffer
\newblock {(Springer, New York, 2007)}.

\bibitem{CHPark1993ZXShen}
Z. X. Shen, D. S. Dessau, B. O. Wells, D. M. King, W. E. Spicer, A. J. Arko, D. Marshall, L. W. Lombardo, A. Kapitulnik, P. Dickinson, S. Doniach, J. DiCarlo, A. G. Loeser, and C. H. Park,
\newblock Anomalously large gap anisotropy in the a-b plane of Bi$_2$Sr$_2$CaCu$_2$O$_{8+\delta}$,
\newblock {Phys. Rev. Lett.} \textbf{70}, 1553 (1993).

\bibitem{HDing1996}
H. Ding, M. R. Norman, J. C. Campuzano, M. Randeria, A. F. Bellman, T. Yokoya, T. Takahashi, T. Mochiku, and K. Kadowaki,
\newblock Angle-resolved photoemission spectroscopy study of the superconducting gap anisotropy in Bi$_2$Sr$_2$CaCu$_2$O$_{8+x}$,
\newblock {Phys. Rev. B} \textbf{54}, R9678 (1996).

\bibitem{ZXShen2014MHashimoto}
M. Hashimoto, E. A. Nowadnick, R. H. He, I. M. Vishik, B. Moritz, Y. He, K. Tanaka, R. G. Moore, D. H. Lu, Y. Yoshida, M. Ishikado, T. Sasagawa, K. Fujita, S. Ishida, S. Uchida, H. Eisaki, Z. Hussain, T. P. Devereaux, and Z. X. Shen,
\newblock Direct spectroscopic evidence for phase competition between the pseudogap and superconductivity in Bi$_2$Sr$_2$CaCu$_2$O$_{8+\delta}$,
\newblock {Nat. Mater.} \textbf{14}, 37 (2015).

\bibitem{ZXShen1996DSMarshall}
D. S. Marshall, D. S. Dessau, A. G. Loeser, C-H. Park, A. Y. Matsuura, J. N. Eckstein, I. Bozovic, P. Fournier, A. Kapitulnik, W. E. Spicer, and Z. X. Shen,
\newblock Unconventional electronic structure evolution with hole doping in Bi$_2$Sr$_2$CaCu$_2$O$_{8+\delta}$: angle-resolved photoemission results,
\newblock {Phys. Rev. Lett.} \textbf{76}, 4841 (1996).

\bibitem{KKapitulnik1996AGLoeser}
A. G. Loeser, Z. X. Shen, D. S. Dessau, D. S. Marshall, C. H. Park, P. Fournier, and A. Kapitulnik,
\newblock Excitation gap in the normal state of underdoped Bi$_2$Sr$_2$CaCu$_2$O$_{8+\delta}$,
\newblock {Science} \textbf{273}, 325 (1996).

\bibitem{JGiapintzakis1996HDing}
H. Ding, T. Yokoya, J. C. Campuzano, T. Takahashi, M. Randeria, M. R. Norman, T. Mochiku, K. Kadowaki, and J. Giapintzakis,
\newblock Spectroscopic evidence for a pseudogap in the normal state of underdoped high-$T_C$ superconductors,
\newblock {Nature} \textbf{382}, 51 (1996).

\bibitem{ZXShen2018IMVishik}
I. M. Vishik,
\newblock Photoemission perspective on pseudogap, superconducting fluctuations, and charge order in cuprates: a review of recent progress,
\newblock {Rep. Prog. Phys.} \textbf{81}, 062501 (2018).

\bibitem{ZXShen2000PVBogdanov}
P. V. Bogdanov, A. Lanzara, S. A. Kellar, X. J. Zhou, E. D. Lu, W. J. Zheng, G. Gu, J. I. Shimoyama, K. Kishio, H. Ikeda, R. Yoshizaki, Z. Hussain, and Z. X. Shen,
\newblock Evidence for an energy scale for quasiparticle dispersion in Bi$_2$Sr$_2$CaCu$_2$O$_8$,
\newblock {Phys. Rev. Lett.} \textbf{85}, 2581 (2000).

\bibitem{DGHinks2001PDJohnson}
P. D. Johnson, T. Valla, A. V. Fedorov, Z. Yusof, B. O. Wells, Q. Li, A. R. Moodenbaugh, G. D. Gu, N. Koshizuka, C. Kendziora, Sha Jian, and D. G. Hinks,
\newblock Doping and temperature dependence of the mass enhancement observed in the cuprate Bi$_2$Sr$_2$CaCu$_2$O$_{8+\delta}$,
\newblock {Phys. Rev. Lett.} \textbf{87}, 177007 (2001).

\bibitem{KKadowaki2001AKaminski}
A. Kaminski, M. Randeria, J. C. Campuzano, M. R. Norman, H. Fretwell, J. Mesot, T. Sato, T. Takahashi, and K. Kadowaki,
\newblock Renormalization of spectral line shape and dispersion below $T_C$ in Bi$_2$Sr$_2$CaCu$_2$O$_{8+\delta}$,
\newblock {Phys. Rev. Lett.} \textbf{86}, 1070 (2001).

\bibitem{ZXShen2001ALanzara}
A. Lanzara, P. V. Bogdanov, X. J. Zhou, S. A. Kellar, D. L. Feng, E. D. Lu, T. Yoshida, H. Eisaki, A. Fujimori, K. Kishio, J. I. Shimoyama, T. Noda, S. Uchida, Z. Hussain, and Z. X. Shen,
\newblock Evidence for ubiquitous strong electron-phonon coupling in high-temperature superconductors,
\newblock {Nature} \textbf{412}, 510 (2001).

\bibitem{ZXShen2003XJZhou}
X. J. Zhou, T. Yoshida, A. Lanzara, P. V. Bogdanov, S. A. Kellar, K. M. Shen, W. L. Yang, F. Ronning, T. Sasagawa, T. Kakeshita, T. Noda, H. Eisaki, S. Uchida, C. T. Lin, F. Zhou, J. W. Xiong, W. X. Ti, Z. X. Zhao, A. Fujimori, Z. Hussain, and Z. X. Shen,
\newblock Universal nodal fermi velocity,
\newblock {Nature} \textbf{423}, 398 (2003).

\bibitem{DSDessau2003ADGromko}
A. D. Gromko, A. V. Fedorov, Y. D. Chuang, J. D. Koralek, Y. Aiura, Y. Yamaguchi, K. Oka, Y. Ando, and D. S. Dessau,
\newblock Mass-renormalized electronic excitations at ($\pi$,0) in the superconducting state of Bi$_2$Sr$_2$CaCu$_2$O$_{8+\delta}$,
\newblock {Phys. Rev. B} \textbf{68}, 174520 (2003).

\bibitem{JFink2003TKKim}
T. K. Kim, A. A. Kordyuk, S. V. Borisenko, A. Koitzsch, M. Knupfer, H. Berger, and J. Fink,
\newblock Doping dependence of the mass enhancement in (Pb,Bi)$_2$Sr$_2$CaCu$_2$O$_8$ at the antinodal point in the superconducting and normal states,
\newblock {Phys. Rev. Lett.} \textbf{91}, 167002 (2003).

\bibitem{ZXShen2004TCuk}
T. Cuk, F. Baumberger, D. H. Lu, N. Ingle, X. J. Zhou, H. Eisaki, N. Kaneko, Z. Hussain, T. P. Devereaux, N. Nagaosa, and Z. X. Shen,
\newblock Coupling of the $B_{1g}$ phonon to the antinodal electronic states of Bi$_2$Sr$_2$Ca$_{0.92}$Y$_{0.08}$Cu$_2$O$_{8+\delta}$,
\newblock {Phys. Rev. Lett.} \textbf{93}, 117003 (2004).

\bibitem{XJZhou2008WTZhang}
W. T. Zhang, G. D. Liu, L. Zhao, H. Y. Liu, J. Q. Meng, X. L. Dong, W. Lu, J. S. Wen, Z. J. Xu, G. D. Gu, T. Sasagawa, G. L. Wang, Y. Zhu, H. B. Zhang, Y. Zhou, X. Y. Wang, Z. X. Zhao, C. T. Chen, Z. Y. Xu, and X. J. Zhou,
\newblock Identification of a new form of electron coupling in the Bi$_2$Sr$_2$CaCu$_2$O$_8$ superconductor by laser-based angle-resolved photoemission spectroscopy,
\newblock {Phys. Rev. Lett.} \textbf{100}, 107002 (2008).

\bibitem{XJZhou2013JFHe}
J. F. He, W. T. Zhang, J. M. Bok, D. X. Mou, L. Zhao, Y. Y. Peng, S. L. He, G. D. Liu, X. L. Dong, J. Zhang, J. S. Wen, Z. J. Xu, G. D. Gu, X. Y. Wang, Q. J. Peng, Z. M. Wang, S. J. Zhang, F. Yang, C. T. Chen, Z. Y. Xu, H. Y. Choi, C. M. Varma, and X. J. Zhou,
\newblock Coexistence of two sharp-mode couplings and their unusual momentum dependence in the superconducting state of Bi$_2$Sr$_2$CaCu$_2$O$_{8+\delta}$ revealed by laser-based angle-resolved photoemission,
\newblock {Phys. Rev. Lett.} \textbf{111}, 107005 (2013).

\bibitem{XJZhou2016JMBok}
J. M. Bok, J. J. Bae, H. Y. Choi, C. M. Varma, W. T. Zhang, J. F. He, Y. X. Zhang, L. Yu, and X. J. Zhou,
\newblock Quantitative determination of pairing interactions for high-temperature superconductivity in cuprates,
\newblock {Sci. Adv.} \textbf{2}, e1501329 (2016).

\bibitem{ZXShen2018YHe}
Y. He, M. Hashimoto, D. Song, S. D. Chen, J. He, I. M. Vishik, B. Moritz, D. H. Lee, N. Nagaosa, J. Zaanen, T. P. Devereaux, Y. Yoshida, H. Eisaki, D. H. Lu, and Z. X. Shen,
\newblock Rapid change of superconductivity and electron-phonon coupling through critical doping in Bi-2212,
\newblock {Science} \textbf{362}, 62 (2018).

\bibitem{JCDavis2001SHPan}
S. H. Pan, J. P. O'Neal, R. L. Badzey, C. Chamon, H. Ding, J. R. Engelbrecht, Z. Wang, H. Eisaki, S. Uchida, A. K. Gupta, K. W. Ng, E. W. Hudson, K. M. Lang, and J. C. Davis,
\newblock Microscopic electronic inhomogeneity in the high-$T_C$ superconductor Bi$_2$Sr$_2$CaCu$_2$O$_{8+x}$,
\newblock {Nature} \textbf{413}, 282 (2001).

\bibitem{JCDavis2002JHoffman}
J. E. Hoffman, E. W. Hudson, K. M. Lang, V. Madhavan, H. Eisaki, S. Uchida, and J. C. Davis,
\newblock A four unit cell periodic pattern of quasi-particle states surrounding vortex cores in Bi$_2$Sr$_2$CaCu$_2$O$_{8+\delta}$,
\newblock {Science} \textbf{295}, 466 (2002).

\bibitem{JCDavis2007Yohsaka}
Y. Kohsaka, C. Taylor, K. Fujita, A. Schmidt, C. Lupien, T. Hanaguri, M. Azuma, M. Takano, H. Eisaki, H. Takagi, S. Uchida, and J. C. Davis,
\newblock An intrinsic bond-centered electronic glass with unidirectional domains in underdoped cuprates,
\newblock {Science} \textbf{315}, 1380 (2007).

\bibitem{JCDavis2016MHamidian}
M. H. Hamidian, S. D. Edkins, S. H. Joo, A. Kostin, H. Eisaki, S. Uchida, M. J. Lawler, E. A. Kim, A. P. Mackenzie, K. Fujita, J. Lee, and J. C. S. Davis,
\newblock Detection of a cooper-pair density wave in Bi$_2$Sr$_2$CaCu$_2$O$_{8+x}$,
\newblock {Nature} \textbf{532}, 343 (2016).

\bibitem{OFisher2007}
Ø. Fischer, M. Kugler, I. M. Aprile, C. Berthod, and C. Renner,
\newblock Scanning tunneling spectroscopy of high-temperature superconductors,
\newblock {Rev. Mod. Phys.} \textbf{79}, 353 (2007).

\bibitem{AWSLEIGHT1988YGao}
Y. Gao, P. Lee, P. Coppens, M. A. Subramanian, and A. W. Sleight,
\newblock The incommensurate modulation of the 2212 Bi-Sr-Ca-Cu-O superconductor,
\newblock {Science} \textbf{241}, 954 (1988).

\bibitem{BGHyde1988RLWithers}
R. L. Withers, J. G. Thompson, L. R. Wallenberg, J. D. FitzGerald, J. S. Anderson, and B. G. Hyde,
\newblock A transmission electron microscope and group theoretical study of the new Bi-based high-$T_C$ superconductors and some closely related Aurivillius phases,
\newblock {J. Phys. C} \textbf{21}, 6067 (1988).

\bibitem{OEibl1991}
O. Eibl,
\newblock Displacive modulation and chemical composition of (Bi, Pb)$_2$Sr$_2$Ca$_{n-1}$Cu$_n$O$_{2n+4}$ ($n$=2, 3) high-$T_C$ superconductors,
\newblock {Physica C} \textbf{175}, 419 (1991).

\bibitem{LJGauckler1994HHeinrich}
H. Heinrich, G. Kostorz, B. Heeb, and L. J. Gauckler, 
\newblock Modelling the atomic displacements in Bi$_2$Sr$_2$Ca$_{n-1}$Cu$_n$O$_x$ superconductors,
\newblock {Physica C} \textbf{224}, 133 (1994).

\bibitem{KKadowaki1994PAebi}
P. Aebi, J. Osterwalder, P. Schwaller, L. Schlapbach, M. Shimoda, T. Mochiku, and K. Kadowaki,
\newblock Complete fermi surface mapping of Bi$_2$Sr$_2$CaCu$_2$O$_{8+x}$(001): Coexistence of short range antiferromagnetic correlations and metallicity in the same phase,
\newblock {Phys. Rev. Lett.} \textbf{72}, 2757 (1994).

\bibitem{KKadowaki1995JOsterwalder}
J. Osterwalder, P. Aebi, P. Schwaller, L. Schlapbach, M. Shimoda, T. Mochiku, K. Kadowaki,
\newblock Angle-resolved photoemission experiments on Bi$_2$Sr$_2$CaCu$_2$O$_{8+\delta}$ (001) Effects of the incommensurate lattice modulation,
\newblock {Appl. Phys. A} \textbf{60}, 247 (1995).

\bibitem{GPBrivio1996HDing}
H. Ding, A. F. Bellman, J. C. Campuzano, M. Randeria, M. R. Norman, T. Yokoya, T. Takahashi, H. K. Yoshida, T. Mochiku, K. Kadowaki, G. Jennings, and G. P. Brivio,
\newblock Electronic excitations in Bi$_2$Sr$_2$CaCu$_2$O$_8$: Fermi surface, dispersion, and absence of bilayer splitting,
\newblock {Phys. Rev. Lett.} \textbf{76}, 1533 (1996).

\bibitem{KKadowaki2000HMFretwell}
H. M. Fretwell, A. Kaminski, J. Mesot, J. C. Campuzano, M. R. Norman, M. Randeria, T. Sato, R. Gatt, T. Takahashi, and K. Kadowaki,
\newblock Fermi surface of Bi$_2$Sr$_2$CaCu$_2$O$_8$,
\newblock {Phys. Rev. Lett.} \textbf{84}, 4449 (2000).

\bibitem{XJZhou2019JLiu}
J. Liu, L. Zhao, Q. Gao, P. Ai, L. Zhang, T. Xie, J. W. Huang, Y. Ding, C. Hu, H. T. Yan, C. Y. Song, Y. Xu, C. Li, Y. Q. Cai, H. T. Rong, D. S. Wu, G. D. Liu, Q. Y. Wang, Y. Huang, F. F. Zhang, F. Yang, Q. J. Peng, S. L. Li, H. X. Yang, J. Q. Li, Z. Y. Xu, and X. J. Zhou,
\newblock Evolution of incommensurate superstructure and electronic structure with Pb substitution in (Bi$_{2-x}$Pb$_x$)Sr$_2$CaCu$_2$O$_{8+\delta}$ superconductors,
\newblock {Chin. Phys. B} \textbf{28}, 077403 (2019).

\bibitem{ZXShen2001PVBogdanov}
P. V. Bogdanov, A. Lanzara, X. J. Zhou, S. A. Kellar, D. L. Feng, E. D. Liu, H. Eisaki, J. I. Shimoyama, K. Kishio, Z. Hussain, and Z. X. Shen,
\newblock Photoemission study of Pb doped Bi$_2$Sr$_2$CaCu$_2$O$_8$: A fermi surface picture,
\newblock {Phys. Rev. B} \textbf{64}, 180505(R) (2001).

\bibitem{ZXShen2001DLFeng}
D. L. Feng, N. P. Armitage, D. H. Lu, A. Damascelli, J. P. Hu, P. Bogdanov, A. Lanzara, F. Ronning, K. M. Shen, H. Eisaki, C. Kim, Z. X. Shen, J. -i. Shimoyama, and K. Kishio,
\newblock Bilayer splitting in the electronic structure of heavily overdoped Bi$_2$Sr$_2$CaCu$_2$O$_{8+\delta}$,
\newblock {Phys. Rev. Lett.} \textbf{86}, 5550 (2001).

\bibitem{DSDessau2001YDChuang}
Y. D. Chuang, A. D. Gromko, A. Fedorov, Y. Aiura, K. Oka, Y. Ando, H. Eisaki, S. I. Uchida, and D. S. Dessau,
\newblock Doubling of the bands in overdoped Bi$_2$Sr$_2$CaCu$_2$O$_{8+x}$: evidence for $c$-axis bilayer coupling,
\newblock {Phys. Rev. Lett.} \textbf{87}, 117002 (2001).

\bibitem{RFollath2002SVBorisenko}
S. V. Borisenko, A. A. Kordyuk, T. K. Kim, S. Legner, K. A. Nenkov, M. Knupfer, M. S. Golden, J. Fink, H. Berger, and R. Follath,
\newblock Superconducting gap in the presence of bilayer splitting in underdoped (Pb,Bi)$_2$Sr$_2$CaCu$_2$O$_{8+\delta}$,
\newblock {Phys. Rev. B} \textbf{66}, 140509(R) (2002).

\bibitem{TXiang2003YHSu}
Y. H. Su , J. Chang, H. T. Lu, H. G. Luo, and T. Xiang,
\newblock Effect of bilayer coupling on tunneling conductance of double-layer high-$T_C$ cuprates,
\newblock {Phys. Rev. B} \textbf{68}, 212501 (2003).

\bibitem{STurchini2004SVBorisenko}
S. V. Borisenko, A. A. Kordyuk, S. Legner, T. K. Kim, M. Knupfer, C. M. Schneider, J. Fink, M. S. Golden, M. Sing, R. Claessen, A. Yaresko, H. Berger, C. Grazioli, and S. Turchini,
\newblock Circular dichroism and bilayer splitting in the normal state of underdoped (Pb,Bi)$_2$Sr$_2$(Ca$_{1-x}$Y$_x$)Cu$_2$O$_{8+\delta}$ and overdoped (Pb,Bi)$_2$Sr$_2$CaCu$_2$O$_{8+\delta}$,
\newblock {Phys. Rev. B} \textbf{69}, 224509 (2004).

\bibitem{DSDessau2004YDChuang}
Y. D. Chuang, A. D. Gromko, A. V. Fedorov, Y. Aiura, K. Oka, Y. Ando, M. Lindroos, R. S. Markiewicz, A. Bansil, and D. S. Dessau,
\newblock Bilayer splitting and coherence effects in optimal and underdoped Bi$_2$Sr$_2$CaCu$_2$O$_{8+\delta}$,
\newblock {Phys. Rev. B} \textbf{69}, 094515 (2004).

\bibitem{YAndo2004AAKordyuk}
A. A. Kordyuk, S. V. Borisenko, A. N. Yaresko, S. L. Drechsler, H. Rosner, T. K. Kim, A. Koitzsch, K. A. Nenkov, M. Knupfer, J. Fink, R. Follath, H. Berger, B. Keimer, S. Ono, and Y. Ando,
\newblock Evidence for CuO conducting band splitting in the nodal direction of Bi$_2$Sr$_2$CaCu$_2$O$_{8+\delta}$,
\newblock {Phys. Rev. B} \textbf{70}, 214525 (2004).

\bibitem{XJZhou2019PAi}
P. Ai, Q. Gao, J. Liu, Y. X. Zhang, C. Li, J. W. Huang, C. Y. Song, H. T. Yan, L. Zhao, G. D. Liu, G. D. Gu, F. F. Zhang, F. Yang, Q. J. Peng, Z. Y. Xu, and X. J. Zhou,
\newblock Distinct superconducting gap on two bilayer-split fermi surface sheets in Bi$_2$Sr$_2$CaCu$_2$O$_{8+\delta}$ superconductor,
\newblock {Chin. Phys. Lett.} \textbf{36}, 067402 (2019).

\bibitem{XJZhou2008GDLiu}
G. D. Liu, G. L. Wang, Y. Zhu, H. B. Zhang, G. C. Zhang, X. Y. Wang, Y. Zhou, W. T. Zhang, H. Y. Liu, L. Zhao, J. Q. Meng, X. L. Dong, C. T. Chen, Z. Y. Xu, and X. J. Zhou,
\newblock Development of a vacuum ultraviolet laser-based angle-resolved photoemission system with a superhigh energy resolution better than 1 meV,
\newblock {Rev. Sci. Instrum.} \textbf{79}, 023105 (2018).

\bibitem{XJZhou2018}
X. J. Zhou, S. L. He, G. D. Liu, L. Zhao, L. Yu, and W. T. Zhang,
\newblock New developments in laser-based photoemission spectroscopy and its scientific applications: a key issues review,
\newblock {Rep. Prog. Phys.} \textbf{81}, 062101 (2018).

\bibitem{XJZhou2009JQMeng}
J. Q. Meng, G. D. Liu, W. T. Zhang, L. Zhao, H. Y. Liu, W. Lu, X. L. Dong, and X. J. Zhou,
\newblock Growth, characterization and physical properties of high-quality large single crystals of Bi$_2$(Sr$_{2-x}$La$_x$)CuO$_{6+\delta}$ high-temperature superconductors,
\newblock {Supercond. Sci. Technol.} \textbf{22}, 045010 (2009).

\bibitem{XJZhou2016YXZhang}
Y. X. Zhang, L. Zhao, G. D. Gu, X. J. Zhou,
\newblock A reproducible approach of preparing high-quality overdoped Bi$_2$Sr$_2$CaCu$_2$O$_{8+\delta}$ single crystals by oxygen annealing and quenching method,
\newblock {Chin. Phys. Lett.} \textbf{33}, 067403 (2016).

\bibitem{TKondo2017SKunisada}
S. Kunisada, S. Adachi, S. Sakai, N. Sasaki, M. Nokayama, S. Akebi, K. Kuroda, T. Sasagawa, T. Watanabe, S. Shin, and T. Kondo,
\newblock Observation of bogoliubov band hybridization in the optimally doped trilayer Bi$_2$Sr$_2$Ca$_2$Cu$_3$O$_{10+\delta}$,
\newblock {Phys. Rev. Lett.} \textbf{119}, 217001 (2017).

\bibitem{ZXShen2000DLFeng}
D. L. Feng, D. H. Lu, K. M. Shen, C. Kim, H. Eisaki, A. Damascelli, R. Yoshizaki, J. i. Shimoyama, K. Kishio, G. D. Gu, S. Oh, A. Andrus, J. O'Donnell, J. N. Eckstein, and Z. X. Shen,
\newblock Signature of superfluid density in the single-particle excitation spectrum of Bi$_2$Sr$_2$CaCu$_2$O$_{8+\delta}$,
\newblock {Science} \textbf{289}, 277 (2000).

\bibitem{ZXShen2004XJZhou}
X. J. Zhou, T. Yoshida, D. H. Lee, W. L. Yang, V. Brouet, F. Zhou, W. X. Ti, J. W. Xiong, Z. X. Zhao, T. Sasagawa, T. Kakeshita, H. Eisaki, S. Uchida, A. Fujimori, Z. Hussain, and Z. X. Shen,
\newblock Dichotomy between nodal and antinodal quasiparticles in underdoped (La$_{2-x}$Sr$_x$)CuO$_4$ superconductors,
\newblock {Phys. Rev. Lett.} \textbf{92}, 187001 (2004).

\bibitem{ZXShen2005KMShen}
K. M. Shen, F. Ronning, D. H. Lu, F. Baumberger, N. J. C. Ingle, W. S. Lee, W. Meevasana, Y. Kohsaka, M. Azuma, M. Takano, H. Takagi, and Z. X. Shen,
\newblock Nodal quasiparticles and antinodal charge ordering in Ca$_{2-x}$Na$_x$CuO$_2$Cl$_2$,
\newblock {Science} \textbf{307}, 901 (2005).

\bibitem{ZXShen2006KTanaka}
K. Tanaka, W. S. Lee, D. H. Lu, A. Fujimori, T. Fujii, Risdiana, I. Terasaki, D. J. Scalapino, T. P. Devereaux, Z. Hussain, and Z. X. Shen,
\newblock Distinct fermi-momentum-dependent energy gaps in deeply underdoped Bi2212,
\newblock {Science} \textbf{314}, 1910 (2006).

\bibitem{ALanzara2018KGotlieb}
K. Gotlieb, C. Y. Lin, M. Serbyn, W. T. Zhang, C. L. Smallwood, C. Jozwiak, H. Eisaki, Z. Hussain, A. Vishwanath, and A. Lanzara,
\newblock Revealing hidden spin-momentum locking in a high-temperature cuprate superconductor,
\newblock {Science} \textbf{362}, 1271 (2018).

\bibitem{JQLi2017CGuo}
C. Guo, H. F. Tian, H. X. Yang, B. Zhang, K. Sun, X. Sun, Y. Y. Peng, X. J. Zhou, and J. Q. Li,
\newblock Direct visualization of soliton stripes in the CuO2 plane and oxygen interstitials in Bi$_2$(Sr$_{2-x}$La$_x$)CuO$_{6+\delta}$ superconductors, 
\newblock {Phys. Rev. Mater.} \textbf{1}, 064802 (2017).

\end{thebibliography}

\newpage

\end{document}